\documentstyle[prb,aps,epsf]{revtex}
\begin{document}

\title{Resonant states and order-parameter suppression \\ 
near point-like impurities in d-wave superconductors}

\author { Alexander Shnirman,    
          \.{I}nanc Adagideli, 
          Paul M.~Goldbart, and
          Ali Yazdani
        }

\address{
Department of Physics and 
Material Research Laboratory, \\  
University of Illinois at Urbana-Champaign,
Urbana, Illinois 61801, U.S.A. \\
}

\maketitle

\begin{abstract}

We examine the role of order-parameter suppression in the development 
of low-energy peaks (i.e., resonances) in the tunneling density of 
states near a non-magnetic impurity in a d-wave 
superconductor. Without order-parameter suppression, the zero-energy 
resonance appears only in the unitary (i.e., strong impurity) limit. 
However, suppression makes the resonance appear even when the impurity 
is much weaker. To model this situation, 
we make the physical hypothesis that the order parameter is reduced 
whenever one electron of a Cooper pair encounters the impurity, a 
hypothesis that retains the exact solvability of the problem.  
In this way, we determine that suppression of the order parameter 
drives the effective strength of the impurity towards the unitary limit.
We determine the order-parameter reduction variationally, and show that 
the ratios between the main energy scales---the band width 
and superconducting gap---strongly affect this reduction and, in 
consequence, the position and width of the resonance.

\end{abstract}


\section{Introduction}
The role played by non-magnetic impurities in high-temperature 
superconductors (HTSCs) represents an important element of the 
subject of high-temperature superconductivity.  In contrast with 
the case of conventional superconductors, in which the s-wave 
symmetry of the order parameter tends to weaken the effect of 
such impurities (c.f.~Anderson's theorem\cite{Anderson}), the 
HTSC materials display rich and interesting sensitivity to the 
amount of disorder, even at low disorder-concentrations.  Indeed, 
many physical properties are affected at low temperatures and 
frequencies, the most direct example being the appearance of a 
nonzero density of states (DOS) at the Fermi level \cite{Gorkov_Kalugin}.
One of the reasons for this sensitivity lies in the properties 
of {\it individual\/} non-magnetic impurities in a d-wave 
superconducting host, such impurities giving rise to resonant 
quasiparticle states at sub-gap energies.  The occurrence of these 
states was predicted theoretically by Balatsky, Salkola and co-workers
\cite{Balatsky_Salkola_Rosengren,Salkola_Balatsky_Scalapino_PRL96}, 
and is consistent with recent experimental observations by Yazdani 
et al.\cite{Yazdani}\thinspace\ These states are 
localized near impurity sites, and have finite life-times, due to 
the existence of bulk quasiparticle states into which they may decay.  
Upon neglecting the changes in the order parameter induced by an 
impurity, it was found that, as the strength of the impurity 
increases, the resonances move towards the Fermi level and their 
widths decrease; only in the unitary (i.e., infinitely-strong impurity) 
limit do the resonances reach the Fermi level and become infinitely 
sharp.  The role of the order-parameter changes (i.e., suppression) 
has been analyzed by several groups (see, e.g., 
Refs.~[\onlinecite{Franz_Kallin_Berlinsky,Hettler_Hirschfeld,Chen_Rainer_Sauls,Flatte_Byers}]). In the present Paper we focus on one particular 
effect of order-parameter suppression which has not been reported 
previously: we show that the suppression of the order parameter 
drives (i.e., renormalizes) the effective strength of the impurity 
towards the unitary limit.  Further, we argue that this renormalization 
may be appreciable in the HTSC materials, and may be important for the 
development of a more complete understanding of the low-temperature 
behavior of  the cuprates. Indeed, as was argued 
in Ref.~[\onlinecite{Hirschfeld_Goldenfeld}], 
the observed low-temperature behavior of cuprates is inconsistent 
with the relatively small suppression of the critical temperature, 
unless the impurities are (or at least behave as if they are) 
in the unitary limit.   

\section{Model of Order-Parameter Suppression Near an Impurity}
Consider a point-like impurity in a two-dimensional 
d-wave superconductor. As a fully self-consistent 
treatment of the order-parameter suppression is out of reach, 
we shall proceed by exploring a physically motivated hypothesis for 
the functional form of this suppression. It is commonly assumed 
\cite{Hettler_Hirschfeld,Chen_Rainer_Sauls} that this suppression 
takes the form
\begin{equation}
\label{COMMON_ASSUMPTION} 
\delta\Delta(r,r') \propto f\left({r+r'\over 2}\right) 
\Delta_{0}(r-r') \ ,
\end{equation} 
where $f$ gives the spatial shape of the suppression, 
$\Delta_{0}(r-r')$ is the 
bulk value of the d-wave order parameter, and the impurity is 
located at the origin. This form includes only the d-wave 
pairing channel, and therefore is very convenient. We 
show, however, that Eq.~(\ref{COMMON_ASSUMPTION}) is meaningful 
only for smooth $f$ (varying on length scales much longer than the 
Fermi wave length $k_{F}^{-1}$), and does not describe the 
physical situation at short distances from the impurity. This 
short-distance behavior is important, as it affects the formation of
the scattering resonances. As an extreme example, consider the setting 
of tight-binding electrons moving on a two-dimensional square lattice 
with on-site repulsion 
and nearest-neighbor attraction (i.e., the simplest situation for d-wave 
superconductivity). If we locate an impurity at the origin and wish 
to suppress the order parameter in the four bonds connecting the origin
to its nearest neighbors, we arrive at the following functional 
form of the suppression:    
\begin{equation}
\label{OUR_ASSUMPTION}
\delta\Delta(r,r') = \alpha \, 
\Big(\, \delta(r)+\delta(r')\, \Big)\,  
\Delta_{0}(r-r') \ ,
\end{equation} 
where $\alpha$ is the amplitude of the suppression 
which has the dimensionality of a volume.
This form may also be used for other cases, 
inasmuch as it encodes the idea that the order 
parameter is altered whenever one of the electrons in the Cooper 
pair encounters the impurity.  The Fourier transform of 
Eq.~(\ref{OUR_ASSUMPTION}) reads 
\begin{equation}
\label{OUR_ASSUMPTION_FOURIER}
\delta\Delta(k,k') \equiv
\int dr\,dr'\, \delta\Delta(r,r')\, \exp(ikr + ik'r') =  
\alpha \, 
\Big(\Delta_{0}(k) + \Delta_{0}(-k')\Big) \ .
\end{equation}
As usual for superconductivity, the most important regime  
is the one in which both $k$ and $k'$ are close to the Fermi 
surface. Thus, the assumption (\ref{OUR_ASSUMPTION}) may be relaxed in 
favor of (\ref{OUR_ASSUMPTION_FOURIER}) near the Fermi surface. The form 
(\ref{OUR_ASSUMPTION_FOURIER}) includes pairing channels other 
than d-wave. Let us, e.g., adopt the tight-binding 
shape of the order parameter: 
$\Delta_{0}(k) = \Delta \phi_{\rm d}(k)$, where 
\begin{equation}
\label{WAVE_FUNCTIONS}
\phi_{\rm d/\rm s}(k) \equiv \cos(a p_{x}) \mp \cos(a p_{y}) \ , 
\ 
\end{equation}
in which $a$ is the lattice constant, and the subscripts $d/s$ stand
for the d and extended-s channels.
By introducing the total $\left[ q \equiv k+k'\right]$ and the relative 
$\left[p \equiv (k-k')/2\right]$ momenta of a Cooper pair we arrive at 
\begin{equation}
\label{DELTA_DELTA}
\delta \Delta_{p,q} =
\alpha\Delta\  
\Big(
\phi_{\rm d}(p)\phi_{\rm s}(q/2) + 
\phi_{\rm s}(p)\phi_{\rm d}(q/2)
\Big)
\ .
\end{equation}
Recent numerical self-consistent simulations \cite{Franz_Kallin_Berlinsky}
yield an order-parameter suppression having a form similar
to (\ref{DELTA_DELTA}), with the d-wave contribution having the 
form-factor of the s-wave symmetry and vice versa. 
Note that at small values of $q$
only the d-wave contribution survives, which reconciles 
Eqs.~(\ref{OUR_ASSUMPTION_FOURIER}) and (\ref{COMMON_ASSUMPTION}).
Further support for the choice of the functional form 
(\ref{OUR_ASSUMPTION_FOURIER}) may be provided by
examining the imbalance in the self-consistent equation for
the order parameter with the impurity present and the order
parameter unchanged (i.e., the model considered in Ref. 
\onlinecite{Salkola_Balatsky_Scalapino_PRL96}). The imbalance means
that the gradient of the free energy in the function space of 
$\Delta(k,k')$ is nonzero and the ``direction'' of this gradient
gives the functional form of the linear response of the 
order parameter $\delta\Delta$ to the presence of the impurity.
We find that the ``direction'' of the imbalance is close to 
(\ref{OUR_ASSUMPTION_FOURIER}), and also has the same symmetry as 
(\ref{OUR_ASSUMPTION_FOURIER}).

\section{Density of States near the Impurity}
First, we investigate the local DOS
near the impurity. We employ the standard $T$-matrix 
technique. The system is governed by the Hamiltonian
\begin{equation}
\label{HAMILTONIAN}
\hat H(r,r') = \hat H_{0}(r,r') + \hat \Sigma(r,r') =  
\hat H_{0}(r,r') + U \delta(r)\delta(r') \hat\sigma_{z} - 
\delta\Delta(r,r') \hat\sigma_{x} \ ,
\end{equation}
where $\hat H_{0}$ is the Bogolubov-de Gennes kernel, which describes 
the unperturbed d-wave superconductor, and the term with the 
coefficient $U$ represents the potential scatterer (i.e., impurity), 
hats denote $2\times2$ matrices in the Nambu space. 

The Dyson equation for the full Matsubara Green function 
$\hat G(r,r',i\omega_{n}) \equiv \big(i\omega_{n} - H(r,r')\big)^{-1}$ 
reads
\begin{mathletters}
\begin{eqnarray}
\label{DYSON_FOR_Ga}
\hat G(r,r') &=& \hat G_{0}(r,r') + \int dx_{1}\,dx_{2} 
            \hat G_{0}(r,x_{1})\,\hat \Sigma(x_{1},x_{2})\,\hat G(x_{2},r') 
            \\
\label{DYSON_FOR_Gb}
&=& \hat G_{0}(r,r') + \int dx_{1}\,dx_{2} 
            \hat G_{0}(r,x_{1})\,\hat T(x_{1},x_{2})\,\hat G_{0}(x_{2},r') \ .
\end{eqnarray}
\end{mathletters}\noindent
Our aim is to find the $T$-matrix $\hat T(x_{1},x_{2})$, which would
allow us to calculate, via Eq.~(\ref{DYSON_FOR_Gb}), the full Green function 
and, therefore, the DOS. The Dyson equation for the $T$-matrix
reads:
\begin{equation}
\label{DYSON_FOR_T}
\hat T(x_{1},x_{2}) = \hat \Sigma(x_{1},x_{2}) + \int dy_{1}\,dy_{2} 
\hat \Sigma(x_{1},y_{1})\,\hat G_{0}(y_{1},y_{2})\,\hat T(y_{2},x_{2}) \ .
\end{equation} 
This model is exactly solvable, owing to the fact that the Fourier transform 
of the self-energy $\hat \Sigma(k,k')$ is a degenerate kernel 
(i.e., a sum of factorized functions of $k$ and $k'$):
\begin{equation}
\label{SIGMA_KK}
\hat \Sigma(k,k') = 
U\hat \sigma_{z} - \alpha\ \Big(\Delta_{0}(k)+\Delta_{0}(-k')\Big) 
\hat \sigma_{x} \ . 
\end{equation}
By using this degeneracy property  we may rewrite Eq.~(\ref{DYSON_FOR_T}) as 
\begin{mathletters}
\begin{eqnarray}
\label{T_FH_EQUATION}
\hat T(k,k') &=& U \hat \sigma_{z} - 
          \alpha\big(\Delta_{0}(k)+\Delta_{0}(k')\big)\, \hat \sigma_{x} + 
          U\,\hat \sigma_{z}\, \hat F(k') -
          \alpha\, \hat \sigma_{x}\, \hat H(k') -
          \alpha\, \Delta_{0}(k)\, \hat \sigma_{x}\, \hat F(k') \ ,
\\
\label{F_DEFINITION}
\hat F(k') &\equiv& V^{-1} \sum_{k''} \hat G_{0}(k'')\, \hat T(k'',k') \ , 
\\
\label{H_DEFINITION}
\hat H(k') &\equiv& V^{-1} \sum_{k''} 
\Delta_{0}(k'') \hat G_{0}(k'')\, \hat T(k'',k') \ , 
\end{eqnarray}
\end{mathletters}\noindent
the remaining task being to determine the as-yet unknown matrix-valued
functions $F$ and $H$.
Here and subsequently, we use the symmetry 
$\Delta_{0}(-k) = \Delta_{0}(k)$. By multiplying Eq.~(\ref{T_FH_EQUATION}) 
first by $\hat G_{0}(k)$ and then by $\Delta_{0}(k) \hat G_{0}(k)$ from the 
left, and integrating over $k$, we obtain a system of linear equations 
for $\hat F$ and $\hat H$:
\begin{mathletters}
\begin{eqnarray}
\label{FH_SYSTEMa}
\big(
U \hat P\, \hat \sigma_{z} - \alpha\, \hat L\, \hat \sigma_{x} - \hat 1
\big)\, \hat F(k) - 
\alpha\, \hat P\, \hat \sigma_{x}\, \hat H(k) &=& 
-U \hat P\, \hat \sigma_{z} + 
\alpha\, \hat L\, \hat \sigma_{x} + 
\alpha\, \hat P\, \Delta_{0}(k)\, \hat \sigma_{x} 
\ , \\
\label{FH_SYSTEMb}
\big(
U \hat L\, \hat \sigma_{z} - \alpha\, \hat M\, \hat \sigma_{x}
\big)\, \hat F(k) -
\big(
\alpha\, \hat L\, \hat \sigma_{x} + \hat 1
\big)\, \hat H(k) &=& 
-U \hat L\, \hat \sigma_{z} + 
\alpha\, \hat M\, \hat \sigma_{x} + 
\alpha\, \hat L\, \Delta_{0}(k)\, \hat \sigma_{x} 
\ ,   
\end{eqnarray}
\end{mathletters}\noindent
where
$\hat P \equiv V^{-1} \sum_{k} \hat G_{0}(k)$, 
$\hat L \equiv V^{-1} \sum_{k} \Delta_{0}(k)\, \hat G_{0}(k)$, and
$\hat M \equiv V^{-1} \sum_{k} \Delta_{0}^{2}(k)\, \hat G_{0}(k)$. 

Thus far, we have not made use of the d-wave symmetry of the order 
parameter, and we have not made any approximation beyond the 
mean-field (Bogolubov-de Gennes) approximation. 
To determine the matrices
$\hat P$, $\hat L$ and $\hat M$ we use the d-wave character of
the order parameter to eliminate integrals of the odd powers
of $\Delta_{0}(k)$, and we assume the presence of particle-hole symmetry 
to eliminate integrals of odd powers of the 
single-electron energy $\epsilon(k)$. Taking for the unperturbed Green 
function 
\begin{equation}
\hat G_{0}(k) = {1\over \Delta_{0}^{2}(k) + \epsilon^{2}(k) + \omega_{n}^2}
\left(
\begin{array}{cc}
-i\omega_{n}-\epsilon(k) & -\Delta_{0}(k) \\
-\Delta_{0}(k) & -i\omega_{n}+\epsilon(k)
\end{array}
\right) 
\end{equation}    
we obtain
$\hat P(i\omega_{n}) = P(i\omega_{n}) (-\hat 1)$, 
$\hat L(i\omega_{n}) = L(i\omega_{n}) (-\hat \sigma_{x})$, 
and $\hat M(i\omega_{n}) = i\omega_{n} L(i\omega_{n}) (-\hat 1)$, where
\begin{mathletters}
\begin{eqnarray}
\label{P_DEFINITION}
P(i\omega_{n}) \equiv V^{-1} \sum_{k}  
{i\omega_{n}\over \Delta_{0}^{2}(k) + \epsilon^{2}(k) + \omega_{n}^2} 
\ ,
\\
\label{L_DEFINITION}
L(i\omega_{n}) \equiv V^{-1} \sum_{k}  
{\Delta_{0}^{2}(k)\over \Delta_{0}^{2}(k) + \epsilon^{2}(k) + \omega_{n}^2} 
\ .
\end{eqnarray}
\end{mathletters}\noindent

The system (\ref{FH_SYSTEMa},\ref{FH_SYSTEMb}) consists of a 
pair of 4-dimensional
systems of linear equations having common coefficient matrices 
and distinct inhomogeneous terms. These two systems may 
be analytically solved, and thus the exact $T$-matrix may be rebuilt using 
Eq.~(\ref{T_FH_EQUATION}). Then the exact Green function can be  
built by using (\ref{DYSON_FOR_Gb}), and hence the local DOS 
may be calculated via
\begin{equation}
\label{NU_DEFINITION}
\nu(r,E) = -{1\over \pi}\, {\rm Im}\, 
\hat G_{1,1}(r,r,E+i\delta) |_{\delta \rightarrow 0}
\ .
\end{equation} 

Before we present exact results for the DOS,
it may be noted that the singular behavior (i.e., resonances) of the 
$T$-matrix at sub-gap energies (for $E<\Delta$) originate only from 
zeros of the determinant $D(E)$ of the system 
(\ref{FH_SYSTEMa},\ref{FH_SYSTEMb}). 
This is so because the right hand side of 
Eqs.~(\ref{FH_SYSTEMa},\ref{FH_SYSTEMb}) has at 
most branch-cuts at the sub-gap energies. Thus, it is instructive 
to write down this determinant:
\begin{eqnarray}
\label{DETERMINANT}
D(E) &=& D_{1}(E)D_{2}(E) \ , \nonumber \\
D_{1,2}(E) &\equiv& 
1-2\alpha L(E) + \alpha^{2} L^{2}(E) - \alpha^{2}EL(E)P(E)
\pm U P(E)  
\ .
\end{eqnarray}
If $\alpha = 0$ (i.e., no order-parameter suppression) 
the characteristic equation 
$D(E)=0$ is identical to the equation for the poles of the $T$-matrix 
obtained in Ref.~[\onlinecite{Salkola_Balatsky_Scalapino_PRL96}]. 
Thus, we may conclude that the role of the order-parameter suppression 
in the present model is to modify (i.e., shift) the resonance found in 
Ref.~[\onlinecite{Salkola_Balatsky_Scalapino_PRL96}], rather than 
to add a new resonance (as results from the model proposed in 
Ref.~[\onlinecite{Hettler_Hirschfeld}]).

We now calculate the DOS  precisely {\it at\/} the impurity. 
We obtain
\begin{equation}
\label{NU_AT_R=0}
\nu(r,E)|_{r=0} 
= 
\nu_0(E) -
{1\over \pi}\, {\rm Im}
\left[
{P(E) \big(\,D_{1}(E) - 1\,\big) \over D_{1}(E)}
\right]
=
{1\over \pi}\, {\rm Im}
\left[
{P(E) \over D_{1}(E)}
\right]
\ ,
\end{equation}
where $\nu_0(E) = (\pi^{-1})\, {\rm Im}\, P(E)$ is the unperturbed DOS 
in the d-wave superconductor. For $\alpha=0$, 
the behavior of $\nu(r,E)$ for $r=0$ was described in 
Ref.~[\onlinecite{Salkola_Balatsky_Scalapino_PRL96}].
There, it was shown that a resonant peak appears at negative energies when 
$\tilde U \equiv U \nu_{F}$ becomes comparable to $1$ ($\nu_{F}$ being the 
DOS at the Fermi surface in the normal state). 
The peak moves toward zero energy and becomes narrower and weaker 
as $\tilde U$ grows. In the unitary limit 
(i.e., for $\tilde U \rightarrow \infty$) the peak disappears. This 
disappearance simply means that an infinitely strong impurity
repels all the electronic density from itself. The resonance 
is still there, and to analyze it, one should 
calculate the tunneling density of states in the vicinity 
of the impurity (see Ref.~[\onlinecite{Salkola_Balatsky_Scalapino_PRL96}]).
There, four maxima appear along the lobes of the d-wave order
parameter at a distance of the order of the Fermi wave length
$\lambda_{F}$ from the impurity site. Moreover, a second resonance,
corresponding to a singularity of the subdeterminant 
$D_{2}$, shows up in the vicinity
of the impurity. The width of this second resonance is exactly equal 
to the width of the first, and the positions of the two 
are symmetric with respect to the Fermi energy. However, the spatial 
density distributions of the two resonances differ from one another. 

The new effect that we report here is that if one fixes the impurity
potential $\tilde U$ and allows the order-parameter suppression 
$\alpha$ to grow instead, the DOS behaves 
similarly to the scenario outlined above. Specifically, the resonances 
move toward zero energy, become sharper, and 
a maximum in the DOS is found along the lobes of $\Delta_{0}(k)$ 
at a distance of the order of $\lambda_{F}$ from the impurity.
Indeed, for $E \rightarrow 0$, we obtain from Eq.~(\ref{L_DEFINITION}) 
that $L(E) \rightarrow 2\Delta\nu_{F}$ [we have taken  
$\Delta_{0}(k) = \Delta \cos(2\phi_{k})$]. In this regime
we can approximate $D_{1,2}(E)$ as $(1-2\tilde\alpha)^2  \pm U P(E)$,
where $\tilde\alpha \equiv \alpha\Delta\nu_{F}$. We find that the positions
($\pm \Omega_{0}$) and the width ($\Gamma$) of the resonances may be 
now determined using the formulas of 
Ref.~[\onlinecite{Salkola_Balatsky_Scalapino_PRL96}]:
\begin{equation}
\label{POSITION_WIDTH}
\Omega_{0} = {\Delta\over 
             2\,\tilde U_{\rm eff}\,{\rm ln}(8 \tilde U_{\rm eff})} 
\ \ \ \ ,\ \ \ \  
\Gamma = {\pi \Omega_{0} \over 2\, {\rm ln}(8 \tilde U_{\rm eff})} \ ,
\end{equation}
in which the original strength of the impurity is substituted 
by a renormalized one, viz.,  
$\tilde U_{\rm eff} \equiv \tilde U / (1-2\tilde\alpha)^2$.
Now, unitary behavior is achieved if
$\tilde U_{\rm eff} \gg 1$, and a strong renormalization of the 
bare strength of the impurity occurs if it is possible for $\tilde \alpha$ 
to be close to $1/2$. In the next section we will argue that such a 
regime may be realistic in the HTSC materials. In the meantime,
let us assume that strong renormalization has occurred,
i.e., that the bare strength of the impurity $\tilde U$ was not large 
enough to cause the unitary behavior but that $U_{\rm eff}$ is.
Then, we may ask the question:
Is there any difference in the spatial distribution of the 
resonant-state density between this case and the case when 
$\tilde U \gg 1$? We find that the only difference is precisely 
{\it at\/} the impurity site: there, the electrons ``know'' that the 
impurity is not so strong, and therefore the DOS
is less suppressed than in the ``true'' unitary limit 
(i.e., $\tilde U \gg 1$). 
Farther from the impurity, however, the two cases are 
indistinguishable. The scaling relation between these two
cases may be expressed as
\begin{equation}
\label{SCALING}
\nu(r,E, \tilde U, \tilde \alpha) \approx
\cases{
{\displaystyle
\nu(r,E, \tilde U_{\rm eff}, \tilde \alpha)\,|_{\tilde \alpha = 0} \over
{\displaystyle (1-2\tilde \alpha)^{2}}}\,,& for $r=0, E \approx 0$; \cr
\noalign{\smallskip} \cr
\nu(r,E, \tilde U_{\rm eff}, \tilde \alpha)\,|_{\tilde \alpha = 0}\,,& for
$|r| > \lambda_{F}, E \approx 0$. \cr 
}
\end{equation}
It is not clear that such a behavior could be 
detectable experimentally because even when STM tips are 
precisely above impurities tunneling occurs over 
some neighborhood of the impurity. However, this is 
at least consistent with the experimental results reported
in Ref.~[\onlinecite{Yazdani}].

\section{Amplitude of the Order-Parameter Suppression}
Our next step is to estimate the amplitude of the 
order-parameter suppression 
$\alpha$. In principle, one may envision two different scenarios. 
In the first, the electron-electron interaction is unchanged by 
the presence of the impurity, and, in the d-wave case, 
the suppression of the order-parameter 
is only due to the pair-breaking effect of the impurity. 
In the second scenario, the 
electron-electron interaction is itself suppressed near 
the impurity, thus furthering the suppression of 
$\Delta(r,r')$. Let us make a very crude estimate for the 
second scenario. We again exploit the tight-binding model, and
assume that the order parameter is zero in the four bonds connecting 
the impurity to its neighbors but unchanged elsewhere.
This regime would be reasonable for purely electronic mechanisms 
of superconductivity, as the local electronic structure is
completely altered by the impurity. 
Then $\alpha \approx a^{d}$ and 
$\tilde \alpha = \alpha\, \Delta\, \nu \approx \Delta/2t$, where 
$2t$ is the band-width and we have used as an estimate 
for the density of states $\nu \approx 1/2t a^{d}$.      
If we take into account the fact that in the definition of
$\tilde \alpha$ the DOS at the Fermi level $\nu_{F}$ should 
be used, whereas the lattice constant $a$ is naturally connected to the 
DOS averaged over the whole band (i.e., $\bar \nu$), we arrive at a more
refined estimate: $\tilde \alpha \approx (\Delta/2t)(\nu_{F}/\bar \nu)$.
We see that in conventional superconductors $\tilde \alpha$ is always small,
and thus essentially no renormalization can happen. 
In the HTSC materials, however,
$\Delta/2t$ can be of the order $0.1$, and the proximity of the van Hove 
singularity makes the factor $\nu_{F}/\bar \nu$ important. Thus,
in this case a strong renormalization situation cannot be ruled out.

Next, we show that, even if the electron-electron interaction 
is unchanged near the impurity, the pair-breaking process creates
a suppression of the order parameter to the value estimated above
[\,$\tilde \alpha \approx (\Delta/2t)(\nu_{F}/\bar \nu)$\,] when 
$\tilde U \approx 1$.
We establish this variationally, i.e., we minimize the 
free energy of the system with respect to $\alpha$. 
To calculate this free energy one has to know the form 
of the electron-electron interaction responsible for 
the superconductivity of the system. 

The most general form of the pairing interaction may be 
written as 
\begin{equation}
\label{EE_INTERACTION}
H_{\rm int} = \ {1\over V^{3}} \sum_{p,p',q,q'}
\ g(p,p';q) \ \delta_{q,q'}  \   
c^{\dag}_{p+q/2,\uparrow} \, c^{\dag}_{-p+q/2,\downarrow} \,
c^{\phantom{\dag}}_{-p'+q'/2,\downarrow} \, 
c^{\phantom{\dag}}_{p'+q'/2,\uparrow} 
\ .
\end{equation}
Note that we have introduced the Kronecker symbol
$\delta_{q,q'}$, corresponding to the conservation of
the total momentum, in order to emphasize the matrix
structure of this interaction. This matrix has
a two-fold index ($p,q$) corresponding to the relative 
and the total momenta of the electrons in a Cooper
pair, respectively. To avoid confusion, we shall use the  
letters $p$ and $q$ for the relative and the total momenta
of the electrons in a Cooper pair, reserving the letter
$k$ for momenta of individual electrons.
The standard Hubbard-Stratonovich  decoupling procedure \cite{Kleinert}
yields the following effective action for the order parameter: 
\begin{equation}
\label{S_EFF}
S = S_{1}+S_{2} \equiv 
\int\nolimits_{0}^{\beta} d\tau 
\left[-
{1\over V^{3}} \sum_{p,p',q,q'}
\bar g (p,p';q) \ \delta_{q,q'} \   
\Delta_{p,q}\,\Delta_{p',q'}^{*}
\right]
-
{\rm Tr}\,{\ln}\, \hat G^{-1} 
\ ,
\end{equation}
where $\hat G^{-1}(r,t;r',t') \equiv 
\hat 1\, \delta(t-t')\,\delta(r-r')\, i\partial/\partial t' - 
\delta(t-t')\,\hat H(r,r')$,
and $\bar g (p,p';q)$ stands for the inverse 
of the $g(p,p';q)$ matrix in $p$ space
[the inversion in $q$ space is trivially
performed in Eq.~(\ref{S_EFF})]:
\begin{equation}
\label{INVERSION}
{1\over V} \sum_{p'} g(p,p';q) \, \bar g (p',p'';q) =
V \delta_{p,p''} \ .
\end{equation}
The electronic free energy $F$ in the mean-field approximation is 
thus given by
\begin{equation}
\label{FREE_ENERGY}
F[\Delta_{p,q}] = {S\over \beta} = F_{1} + F_{2}=
-
{1\over V^{3}} \sum_{p,p',q}
\bar g (p,p';q) \    
\Delta_{p,q}\,\Delta_{p',q}^{*}
-
{1\over \beta} {\rm Tr}\,{\ln}\,G^{-1} 
\ .
\end{equation}

We now minimize (\ref{FREE_ENERGY}) with respect
to $\alpha$. The second term of (\ref{FREE_ENERGY}) is treated readily;
indeed
\begin{equation}
\label{F2_DERIVATIVE}
{\partial\over \partial \alpha} F_{2} = 
-{\partial\over\partial\alpha}
\left(
{1\over \beta}{\rm Tr}\,{\ln}\,\hat G^{-1}
\right) 
= 
-{1\over \beta}{\rm Tr}\,\hat G {\partial\over\partial\alpha}\hat G^{-1}
=
-{1\over \beta V^{2}} \sum_{k,k',\omega_{n}} {\rm tr}
\left[
\hat G(k,k',i\omega_{n})\,(\Delta_{0}(k)+\Delta_{0}(k')\,)\,\hat \sigma_{x}
\right]
\ ,
\end{equation}
where the symbol ${\rm tr}$ stands for the trace in the Nambu space only.
The Green function $\hat G$ is known exactly: 
$\hat G(k,k') = V \delta_{k,-k'} \, \hat G_{0}(k) + 
\hat G_{0}(k)\, \hat T(k,k')\, \hat G_{0}(k')$.
Therefore, we can rewrite Eq.~(\ref{F2_DERIVATIVE}) as
\begin{equation}
\label{F2_DERIVATIVE_FACTORIZED}
{\partial\over \partial \alpha} F_{2} = 
-
{2\over \beta V} \sum_{k,\omega_{n}} 
\Delta_{0}(k)\,
{\rm tr}
\left[
\hat G_{0}(k,i\omega_{n})\,\hat \sigma_{x}
\right] 
-
{1\over \beta V^{2}} \sum_{k,k',\omega_{n}} 
\left[
\Delta_{0}(k)+\Delta_{0}(k')
\right]\,
{\rm tr}
\left[
\hat G_{0}(k,i\omega_{n})\hat T(k,k',i\omega_{n})
\hat G_{0}(k',i\omega_{n}) 
\hat \sigma_{x}
\right]
\ .
\end{equation}

As for the first term of Eq.~(\ref{FREE_ENERGY}), we recall that 
\begin{equation}
\label{ORDER_PARAMETER}
\Delta_{p,q} = V\delta_{q,0}\,\Delta_{0}(p) - 
\alpha \, \big(\Delta_{0}(p+q/2) + \Delta_{0}(p-q/2)\big) 
\ ,
\end{equation} 
and therefore $F_{1}$ is a quadratic polynomial in $\alpha$.
The derivative of $F_{1}$ with respect to $\alpha$ 
contains a term independent of $\alpha$ and a term
linear in $\alpha$. It is straightforward to verify
that the term independent of $\alpha$ in $\partial F_{1}/\partial\alpha$, 
when combined with the first term in Eq.~(\ref{F2_DERIVATIVE_FACTORIZED}),
cancel, as together they constitute the BCS self-consistency equation 
for the unperturbed order-parameter.
(The appearance of this equation was to be expected as 
it emerges from the condition that the free energy be
minimal when no impurity is present.) 
We denote the parts of $F_{1}$ and $F_{2}$ remaining after the 
cancellation as $F_{1}^{(2)}$ and $F_{2}^{(2)}$, respectively, and  
the variational condition for $\alpha$ now reads:
\begin{mathletters}
\begin{eqnarray}
\label{EQUATION_FOR_ALPHA}
{\partial F \over \partial \alpha} &=& 
{\partial\over\partial\alpha} F_{1}^{(2)} 
-
{\partial\over\partial\alpha} F_{2}^{(2)} 
=0
\ ,
\\
\label{F_2_2_DEFINITION}
{\partial\over\partial\alpha} F_{2}^{(2)} &\equiv&
{1\over \beta V^{2}} \sum_{k,k',\omega_{n}} 
2 \Delta_{0}(k)
{\rm tr}
\left[
\hat G_{0}(k,i\omega_{n})\,\hat T(k,k',i\omega_{n})\,
\hat G_{0}(k',i\omega_{n})\, 
\hat \sigma_{x}
\right]
\ ,
\\
\label{F_1_2_DEFINITION}
F_{1}^{(2)} &\equiv&
-{1\over V^{3}} \sum_{p,p',q}
\bar g (p,p';q) \   
\delta\Delta_{p,q}\, \delta\Delta_{p',q}^{*}
\ . 
\end{eqnarray}
\end{mathletters}\noindent
To estimate $F_{1}^{(2)}$, we choose the standard form of the interaction, 
which has no $q$-dependence: 
\begin{equation}
\label{G_KK}
g(p,p';q) = -g_{\rm d}\,\phi_{\rm d}(p)\,\phi_{\rm d}(p')  
            -g_{\rm s}\,\phi_{\rm s}(p)\,\phi_{\rm s}(p') \ .
\end{equation}
We include here both d and s channel interactions.
The part of $F_{1}$ quadratic in $\alpha$ reads 
\begin{equation}
\label{F1_2_NO_CUTOFF}
F_{1}^{(2)} = 
{1\over V^{3}} \sum_{p,p',q} 
\left(
{1\over g_{\rm d}} {\phi_{\rm d}(p)\phi_{\rm d}(p') \over N_{\rm d}^{2}} + 
{1\over g_{\rm s}} {\phi_{\rm s}(p)\phi_{\rm s}(p') \over N_{\rm s}^{2}}
\right) 
\delta\Delta_{p,q}\,\delta\Delta_{p',q}^{*}
\ ,
\end{equation}
where $N_{s,d} \equiv V^{-1} \sum_{p} \phi_{\rm s/d}^{2}(p)$.
  
To proceed further we need to assume some particular form of
the wave functions $\phi_{\rm d/s}$. The simplest choice is the 
tight-binding one [Eqs. (\ref{WAVE_FUNCTIONS}) 
and (\ref{DELTA_DELTA})],
for which $F_{1}^{(2)}$ is readily calculated:
\begin{equation}
\label{F1_2_NO_CUTOFF_FINAL}
F_{1}^{(2)} = {\alpha^{2} \Delta^{2} \over a^{d}} 
\left(g_{\rm d}^{-1} + g_{\rm s}^{-1}\right) \ ,
\end{equation}
where we have used the identity 
${V^{-1}} \sum_{q} \phi_{\rm d/s}^{2}(q/2) = 1/a^{d}$.
The appearance of the volume of the lattice cell $a^{d}$ in 
Eq.~(\ref{F1_2_NO_CUTOFF_FINAL}) introduces the band-width 
energy-scale $2t$, via $\bar \nu \approx 1/2t a^{d}$.
Defining, as usual, the pair of dimensionless 
coupling constants $\tilde g_{\rm d/s} \equiv \nu_{F}\, g_{\rm d/s}$,
and differentiating Eq.~(\ref{F1_2_NO_CUTOFF_FINAL}) with 
respect to $\alpha$, we obtain
\begin{equation}
\label{D_F_D_ALPHA_DIMENSIONLESS}
{\partial  \over \partial \alpha}  F_{1}^{(2)} =
{2 \alpha \Delta^{2} \over a^{d}} 
\big(g_{\rm d}^{-1} + g_{\rm s}^{-1}\big) =
2 \tilde \alpha   
\big(\tilde g_{\rm d}^{-1} + \tilde g_{\rm s}^{-1}\big)
\left({2t \bar\nu \over \Delta\,\nu_{F}}\right)
\Delta^{2}\nu_{F}
\ .
\end{equation}

\begin{figure}  
\epsfysize=30\baselineskip
\centerline{\hbox{\epsffile{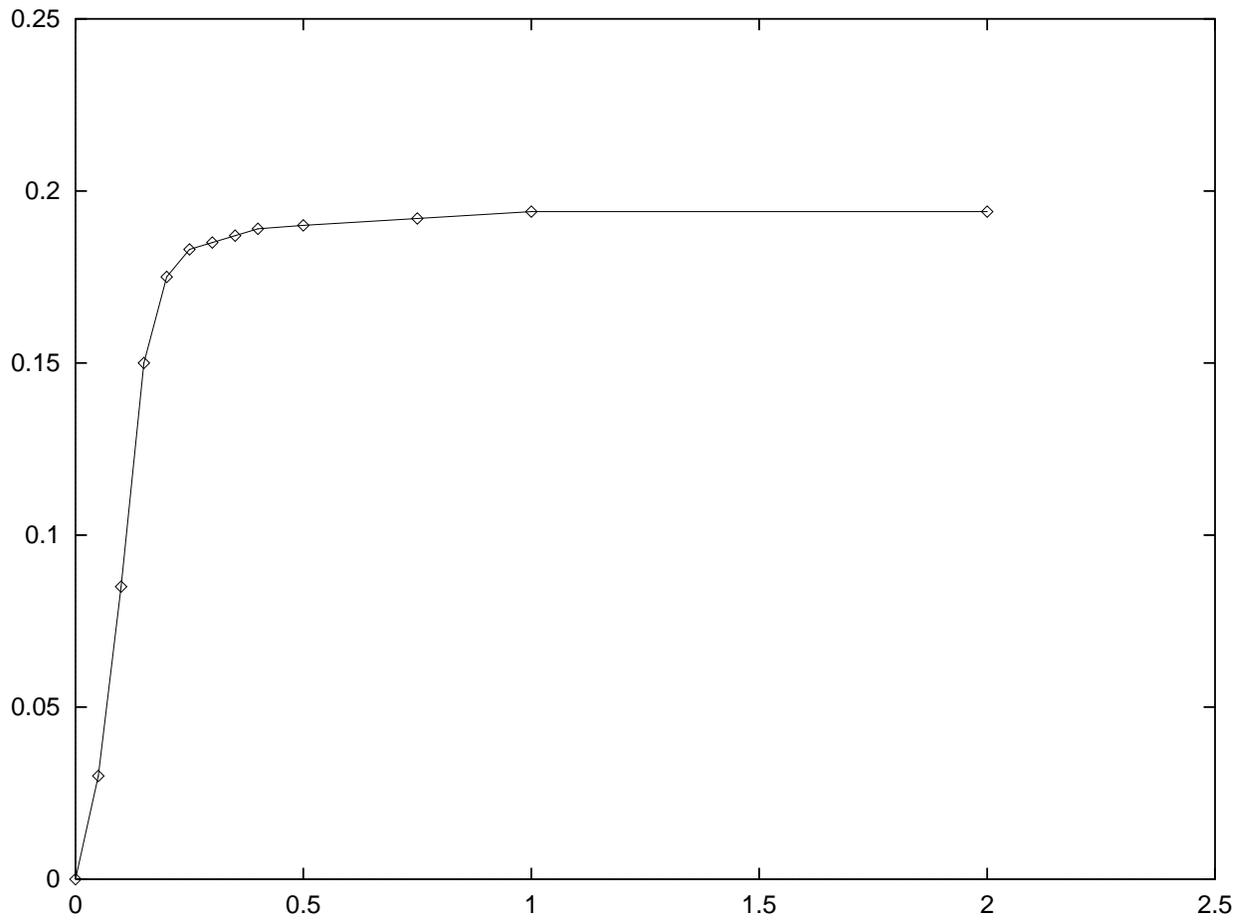}}}
\vskip 0.5cm
\caption[]{\label{ALPHA_U}
Results of the numerical simulations for $\tilde\alpha(\tilde U)$.
$\Delta/2t = 0.1$, $\nu_{F}/\bar \nu = 1$,
the coupling constant $\tilde g_{\rm d} \approx 0.8$ 
is determined self-consistently, and it is assumed that  
$\tilde d_{\rm s} = \tilde d_{\rm d}$.}
\end{figure}

Finally, to find $\tilde \alpha$ we must evaluate 
$\partial F_{2}^{(2)} / \partial\alpha$, as given by 
Eq.~(\ref{F_2_2_DEFINITION}). Although all the components 
of (\ref{F_2_2_DEFINITION}) are known analytically and the 
integrals over $k$ and $k'$ can be expressed via the functions
$P(i\omega_{n})$ and $L(i\omega_{n})$ defined by 
Eqs.~(\ref{P_DEFINITION}) and (\ref{L_DEFINITION}), the remaining 
sum over the Matsubara frequencies must be carried out numerically.
An analytical result is obtained only in the ``true'' unitary
limit $\tilde U \rightarrow \infty$:
\begin{equation}
\label{F_2_ASIMPTOTIC}
{\partial\over\partial\alpha} F_{2}^{(2)}
\rightarrow
{1\over \beta} \sum_{\omega_{n}} 4 L(i\omega_{n}) 
=
4 \tilde g_{\rm d}^{-1} \Delta^{2}\,\nu_{F} \ , 
\end{equation}
where the last equation is obtained from the self-consistency 
condition for the unperturbed order parameter (without impurities).
>From Eqs.~(\ref{D_F_D_ALPHA_DIMENSIONLESS}) and (\ref{F_2_ASIMPTOTIC})
we obtain 
\begin{equation}
\label{ALPHA_ASIMPTOTIC}
\tilde \alpha (\tilde U) |_{\tilde U \rightarrow \infty} = 
{2 \tilde g_{\rm d}^{-1} \over \tilde g_{\rm d}^{-1} + \tilde g_{\rm s}^{-1}}
\ {\Delta\over 2t}\ {\nu_{F}\over\bar \nu} 
\approx 
{\Delta\over 2t}\ {\nu_{F}\over\bar \nu} \ .
\end{equation}
This result would still be meaningless if the asymptotic value 
(\ref{ALPHA_ASIMPTOTIC}) were achieved only for 
$\tilde U \rightarrow \infty$, as the renormalization 
has no effect in the ``true'' unitary limit.
To check how fast this asymptotic value is achieved, we have 
solved Eq.~(\ref{EQUATION_FOR_ALPHA}) numerically for different 
choices of the system parameters. We observe that the asymptotic value is
always reached already for $\tilde U \approx 1$ 
(see, e.g., Fig.~\ref{ALPHA_U}). Thus, impurities having ``mild'' 
values of $\tilde U$ may be renormalized to the unitary limit. 
We were unable, however, to approach numerically the regime 
$\tilde \alpha \approx 1/2$ without employing the van~Hove
singularity (i.e., for $\nu_{F}/\bar \nu \approx 1$). This 
is because the ratio $\Delta/2t$ would have to become 
of order $1$, which is inconsistent with the BCS approximation. 
On the other hand, exploring the van Hove scenario would demand
calculations with realistic band structures 
(see, e.g., Ref.~[\onlinecite{Flatte_Byers}]), the task we leave 
for the future.

\section{Conclusions}
To conclude, we have investigated the role of the order-parameter 
suppression in the development of resonant scattering states 
around impurities in d-wave superconductors. We show that 
the suppression of the order parameter drives the effective 
strength of the impurity towards the unitary limit. This effect 
may be relevant in the HTSC materials, due to the relatively large 
value of the ratio between the superconducting gap $\Delta$ and the 
band-width and due to the possibility of a  van Hove singularity 
in the DOS. The electronic DOS around a renormalized impurity 
is indistinguishable from the DOS around a ``truly'' unitary 
impurity, except precisely at the impurity site. Further 
calculations involving the strong-coupling regime and the 
effects of the real band-structures are needed to establish
the feasibility of the strong renormalization of the 
strength of impurities.   

\acknowledgments
We thank I.~Aleiner, J.~Schmalian, Y.~Lyanda-Geller, 
D.~Maslov, R.~Ramazashvili for fruitful discussions. 
This work was supported by the 
U.S.~Department of Energy, Division of Materials Science, 
under Award No.~DEFG02-96ER45439 through the 
University of Illinois Materials Research Laboratory (AS, \.{I}A, PG)
and by the Fulbright Foundation (AS).



\begin{thebibliography}{10}

\bibitem{Anderson}
P.~W.~Anderson, J. Phys. Chem. Sol. {\bf 11}, 26 (1959).

\bibitem{Gorkov_Kalugin}
L.~P.~Gorkov and P.~A.~Kalugin, JETP Lett. {\bf 41}, 253 (1985).

\bibitem{Balatsky_Salkola_Rosengren}
A.~V.~Balatsky, M.~I.~Salkola, and A.~Rosengren, Phys. Rev. B {\bf 51}, 15547
  (1995).

\bibitem{Salkola_Balatsky_Scalapino_PRL96}
M.~I.~Salkola, A.~V.~Balatsky, and D.~J.~Scalapino, Phys. Rev. Lett. {\bf 77},
  1841 (1996).

\bibitem{Yazdani}
A.~Yazdani et. al., manuscript in preparation (1998).

\bibitem{Franz_Kallin_Berlinsky}
M.~Franz, C.~Kallin, and A.~J.~Berlinsky, Phys. Rev. B {\bf 54}, R6897 (1996).

\bibitem{Hettler_Hirschfeld}
M.~H.~Hettler and P.~J.~Hirschfeld, preprint, cond-mat/9809263.

\bibitem{Chen_Rainer_Sauls}
D.~C.~Chen, D.~Rainer, and J.~A.~Sauls, preprint.

\bibitem{Flatte_Byers}
M.~E.~Flatte and J.~M.~Byers, Phys. Rev. Lett. {\bf 80}, 4546 (1998).

\bibitem{Hirschfeld_Goldenfeld}
P.~J.~Hirschfeld and N.~Goldenfeld, Phys. Rev. B {\bf 48}, 4219 (1993).

\bibitem{Kleinert}
See, e.g., H.~Kleinert, Fortschr. Phys. {\bf 26}, 565 (1978).

\end{thebibliography}
\end{document}